\documentstyle[amstex,amssymb,12pt,epsfig]{article}

\newlength{\dinwidth}
\newlength{\dinmargin}
\begin{document}


\thispagestyle{empty} \vspace*{1cm} \rightline{Napoli DSF-T-4/2008} %
\rightline{INFN-NA-4/2008} \vspace*{2cm}

\begin{center}
{\LARGE Topologically protected qubits as minimal Josephson junction arrays
with non trivial boundary conditions: a proposal}

{\LARGE \ }

{\large Gerardo Cristofano\footnote{{\large {\footnotesize Dipartimento di
Scienze Fisiche,}{\it \ {\footnotesize Universit\'{a} di Napoli ``Federico
II''\ \newline
and INFN, Sezione di Napoli},}{\small Via Cintia, Compl.\ universitario M.
Sant'Angelo, 80126 Napoli, Italy}}}, Vincenzo Marotta\footnote{{\large
{\footnotesize Dipartimento di Scienze Fisiche,}{\it \ {\footnotesize %
Universit\'{a} di Napoli ``Federico II''\ \newline
and INFN, Sezione di Napoli},}{\small Via Cintia, Compl.\ universitario M.
Sant'Angelo, 80126 Napoli, Italy}}},} {\large Adele Naddeo\footnote{{\large
{\footnotesize Dipartimento di Fisica {\it ''}E. R. Caianiello'',}{\it \
{\footnotesize Universit\'{a} degli Studi di Salerno \ \newline
and CNISM, Unit\`{a} di Ricerca di Salerno, }}{\small Via Salvador Allende,
84081 Baronissi (SA), Italy}}}, Giuliano Niccoli\footnote{{\large
{\footnotesize Theoretical Physics Group, DESY, NotkeStra\ss e 85 22603,
Hamburg, Germany.}}}}

{\small \ }

{\bf Abstract\\[0pt]
}
\end{center}

\begin{quotation}
Recently a one-dimensional closed ladder of Josephson junctions
has been studied \cite{noi6} within a twisted conformal field
theory (CFT) approach \cite{cgm2,cgm4} and shown to develop the
phenomenon of flux fractionalization \cite{noi4}. That led us to
predict the emergence of a topological order in such a system
\cite{noi3}. In this letter we analyze the ground states and the
topological properties of fully frustrated Josephson junction
arrays (JJA) arranged in a Corbino disk geometry for a variety of
boundary conditions. \ In particular minimal configurations of
fully frustrated JJA are considered and shown to exhibit the
properties needed in order to build up a solid state qubit,
protected from decoherence. The stability and transformation
properties of the ground states of the JJA under adiabatic
magnetic flux changes are analyzed in detail in order to provide a
tool for the manipulation of the proposed qubit.

\vspace*{0.5cm}


{\footnotesize PACS: 11.25.Hf, 03.75.Lm, 74.81.Fa\newpage }\baselineskip%
=18pt \setcounter{page}{2}
\end{quotation}

\section{Introduction}

Today the physical realization of a quantum computer represents a very hard
task because of the limits imposed by decoherence and because interactions
between qubits cannot be well controlled. A lot of solid state qubit
implementations have been proposed in the last ten years; among them,
systems based on Josephson junctions \cite{revqubit} are promising because
of the well assessed fabrication technology, which could allow a high degree
of scalability. But anyway the path towards the physical realization of a
fully integrated quantum computer still remains a remarkable challenge.
Recently topologically ordered quantum systems have been proposed as
physical analogues of quantum error-correcting codes \cite{shor} starting
from Kitaev seminal work \cite{kitaev}: all such systems share a built in
protection from decoherence. The idea of Kitaev involves a protected
subspace created by a topological degeneracy of the ground state: such a
degeneracy is typically due to a conservation law such as the conservation
of the parity of the number of particles along some long contour. The
underlying idea in Kitaev work is the notion of topological order, that is a
new kind of order which cannot be due to a spontaneous symmetry breakdown.
In such a case the system lacks of local order parameters but displays a
weak form of order, which is sensitive to the topology of the underlying two
dimensional manifold. Such a concept was first introduced in order to
describe the ground state of a quantum Hall fluid \cite{wen} but today it is
of much more general interest \cite{wen1}\cite{nussinov}. Two features of
topological order are very striking: fractionally charged quasiparticles and
a ground state degeneracy depending on the topology of the underlying
manifold, which is lifted by quasiparticles tunneling processes. In general
a system is in a topological phase if its low-energy, long-distance
effective field theory is a topological quantum field theory that is, if all
of its physical correlation functions are topologically invariant up to
corrections of the form $e^{-\frac{\Delta }{T}}$at temperature $T$ for some
nonzero energy gap $\Delta $. More recently superconductors have been
proposed \cite{fisher1} in which superconductivity arises from a topological
mechanism rather than from a Ginzburg-Landau paradigm: the key feature is a
mapping on an effective Chern-Simons gauge theory, which turns out to be
exact in the case of JJA and frustrated JJA \cite{pasquale}. Non-Abelian
quantum Hall states, in particular the $\nu =\frac{5}{2}$ one, appear also
very promising in that the quasiparticle excitations are non-Abelian anyons
\cite{nonabelian} obeying non-Abelian braiding statistics: in such a case
quantum information is codified in states with multiple quasiparticles,
which show a topological degeneracy. Furthermore topological phases have
been recently recognized in chiral $p$-wave superconductors \cite{metaloxide}%
, such as $Sr_{2}RuO_{4}$, and ultra-cold atomic gases \cite{optical1}.
Large and small size Josephson junction arrays of special geometry \cite
{ioffe}\cite{ioffe1} have been proposed as well, which share the property
that, in the classical limit for the local superconducting variables, the
ground state is highly degenerate. The residual quantum processes within
such a low energy subspace lift the classical degeneracy in favor of
macroscopic coherent superpositions of classical ground states \cite{ioffe1}%
. The protected degeneracy in all such systems emerges as a natural property
of the lattice Chern-Simons gauge theories which describe them \cite{ioffe1}%
. In general, if a physical system has topological degrees of freedom that
are insensitive to local perturbations (that is noise), then information
contained in those degrees of freedom would be automatically protected
against errors caused by local interactions with the environment \cite
{kitaev}. The procedure implemented in all such realizations of protected
JJA based qubits runs as follows. A quantum system is required with $2^{K}$\
quantum states ($K$\ being the number of big openings in the Josephson
systems under study) which are degenerate in the absence of external
perturbations and are robust against local random fluctuations. This means
that the Hilbert space should contain a $2^{K}$-dimensional subspace
characterized by the crucial property that any local operator $\widehat{O}$
has only state-independent diagonal matrix elements up to vanishingly small
corrections: $\left\langle n\right| \widehat{O}\left| m\right\rangle
=O_{0}\delta _{mn}+o\left[ \exp \left( -L\right) \right] $, $L$ being the
system size. So a possible answer to such a highly non trivial requirement
could be a system with a protected subspace built up by a topological
degeneracy of the ground state \cite{kitaev}. An alternative procedure could
be to exhibit a low-energy effective field theory for the system under study
which is a topological one and whose vacua are topologically degenerate and
robust against noise. Following such an approach, we were able to predict
the emergence of topological order in a fully frustrated Josephson junction
ladder (JJL) with Mobius boundary conditions \cite{noi3,noi4} and to build
up a protected subspace with $2^{K}$ quantum states, $K=1$, which could be
identified with the two states of a ``protected'' qubit \cite{noi6}. The
low-energy effective field theory which we set up is a twisted conformal
field theory, the Twisted Model (TM) \cite{cgm2,cgm4}: it accounts very well
for the topological properties of the system under study \cite{noi3,noi4}.
This finding is crucial and relies on the well known relation between
Chern-Simons gauge theories in $(2+1)$-dimensions and $2$-dimensional CFT
\cite{witten1}.

The aim of this letter is to analyze the ground states and the
topological properties of fully frustrated Josephson junction
arrays (JJA) arranged in a Corbino disk geometry for a variety of
boundary conditions, employing a twisted CFT approach
\cite{cgm2,cgm4} which has been successfully applied to quantum
Hall systems in the presence of impurities or defects \cite
{noi1,noi2,noi5} and to the study of the phase diagram of the
fully frustrated $XY$ model ($FFXY$) on a square lattice
\cite{noi}. In this way we extend to a more general and more
involved system the results obtained in the simplest JJL case
\cite{noi3,noi4,noi6} and exhibit the minimal configuration and
properties needed in order to build up a solid state qubit,
protected from decoherence. We analyze in detail the stability and
transformation properties of the ground state wave functions under
adiabatic magnetic flux changes in order to provide a tool for the
manipulation of the proposed qubit \cite{qcrew1}.

The letter is organized as follows.

In Section 2 we briefly discuss all the relevant phenomenology of $2$%
-dimensional fully frustrated JJA with an emphasis on the underlying
physical model, the $FFXY$ model on a square lattice \cite{ffxy1}. The
analysis will focus on some models which share the same $U\left( 1\right)
\otimes Z_{2}$ degenerate ground state \cite{foda} and are believed to be in
the same universality class.

In Section 3 we recall some aspects of the $m$-reduction procedure, in
particular we show how the $m=2$, $p=0$ case gives rise to the required $%
U\left( 1\right) \otimes Z_{2}$ mixed symmetry \cite{noi} of the $FFXY$
model. Then we focus on the discrete version of such a procedure. In such a
framework we give the whole primary field content of the theory on the plane.

In Section 4 we properly define a JJA on the torus topology, which is
equivalent to the Corbino disk geometry with coinciding boundary, and
introduce the magnetic translation operators. We build up the ground state
wave functions in order to provide a physical identification of the
characters of our CFT, the TM, as the components of the ``center of charge''
for such wave functions. Then we describe the general topological structure
of the vacua on the torus corresponding to all the possible boundary
conditions.

In Section 5 we study the stability and transformation properties of the
four ground states of the JJA, arranged in the closed geometry under
adiabatic magnetic flux changes through the central hole of the Corbino
disk. That allows us to identify the two states of a possible flux qubit
\cite{orlando1}, protected from decoherence, and to provide a tool for its
manipulation.

In Section 6 some comments and outlooks are given while in the Appendix the
explicit expressions of\ the TM characters on the torus are presented in
detail.

\section{Fully frustrated Josephson junction arrays}

Josephson junction arrays (JJA) are prototype systems to investigate a
variety of phase transitions induced by thermal or quantum fluctuations \cite
{fluct1}. If the superconducting islands are of submicron size, quantum
fluctuations play a crucial role and drive the zero temperature
Superconductor-Insulator (SI) phase transition \cite{si1}. There are two
energy scales in JJA: the Josephson energy $J$, which is associated to the
tunneling of Cooper pairs between neighboring islands, and the charging
energy $U$, which is the energy needed in order to add an extra Cooper pair
on a neutral island. A finite $U$ is responsible of quantum fluctuations of
the phases $\left\{ \varphi \right\} $ of the superconducting order
parameter on each island. In the regime $J\gg U$, named the {\it classical}
case, the fluctuations of the phases are small, the system is globally
coherent and superconducting. Conversely, in the opposite regime $J\ll U$,
strong quantum phase fluctuations prevent the array from reaching long-range
phase coherence and make it a Mott insulator. Magnetic frustration can be
introduced in a JJA by applying a magnetic field transversal to the plane of
the array \cite{teitel}\cite{halsey1}; it can be defined as $f=\frac{\Phi }{%
\Phi _{0}}$, where $\Phi $ is the magnetic flux threading each plaquette and
$\Phi _{0}=\frac{hc}{2e}$ is the superconducting flux quantum. The external
magnetic field induces vortices in the array and, in the case of rational
frustration $f=\frac{p}{q}$, the ground state shows a checkerboard
configuration of vortices on a $q\times q$ elementary supercell. In the case
of full frustration, i.e. $f=\frac{1}{2}$, of interest to us here, there are
two degenerate ground states built of a vortex lattice with a $2\times 2$
elementary supercell; the corresponding current flows either clockwise or
anticlockwise in each plaquette, giving rise to a chiral phase
configuration. Quantum fluctuations affect the superconducting regime but do
not destroy the checkerboard structure of the ground state \cite{stroud}.
All the relevant physics of the JJA is captured by the quantum phase model
(QPM), whose Hamiltonian is:
\begin{equation}
H_{QPM}=\sum_{i,j}\left( n_{i}-n_{x}\right) U_{ij}\left( n_{j}-n_{x}\right)
-J\sum_{\left\langle ij\right\rangle }\cos \left( \varphi _{i}-\varphi
_{j}-A_{ij}\right) ,  \label{qpm1}
\end{equation}
where $n_{i}$ and $\varphi _{i}$ are canonically conjugated variables
defined on the sites and satisfying the commutation relations $\left[
\varphi _{i},n_{j}\right] =2ei\delta _{ij}$, the second sum is over nearest
neighbors, $J>0$ is the Josephson energy, the matrix $%
U_{ij}=4e^{2}C_{ij}^{-1}$ describes the Coulomb interaction ($C_{ij}$ is the
capacitance matrix), the external \ voltage $V_{x}$ enters through the
induced charge $en_{x}$ and fixes the average charge on each island and $%
A_{ij}=\frac{2e}{\hslash c}\int_{i}^{j}A{\cdot }dl$ is the line integral
along the bond between adjacent sites $i$ and $j$. We consider the case
where the bond variables $A_{ij}$ are fixed, uniformly quenched, out of
equilibrium with the site variables and satisfy the condition $%
\sum_{p}A_{ij}=2\pi f$; here the sum is over each set of bonds of an
elementary plaquette and $f$ is the strength of frustration. In the
following we will focus on the limit $J\gg U$, where the JJA in the presence
of an external magnetic field transversal to the lattice plane is a physical
realization of the fully frustrated $XY$ model on the square lattice. Let us
then focus on such a model, which is described by the action:
\begin{equation}
H=-J\sum_{\left\langle ij\right\rangle }\cos \left( \varphi _{i}-\varphi
_{j}-A_{ij}\right) .  \label{act1}
\end{equation}
We assume that the local magnetic field in Eq. (\ref{act1}) is equal to the
uniform applied field; such an approximation is more valid the smaller is
the sample size $L$ compared with the transverse penetration depth $\lambda
_{\perp }$. In the case under study, $f=\frac{1}{2}$, such a model has a
continuous $U(1)$ symmetry associated with the rotation of spins and an
extra discrete $Z_{2}$ symmetry, as it has been shown analyzing the
degeneracy of the ground state \cite{ffxy1}. Choosing the Landau gauge, such
that the vector potential vanishes on all horizontal bonds and on
alternating vertical bonds, we get a lattice where each plaquette displays
one antiferromagnetic and three ferromagnetic bonds. Such a choice
corresponds to switching the sign of the interaction term in Eq. (\ref{act1}%
) and is closely related to the presence of two ground states with opposite
chiralities, the first one invariant under shifts by two lattice spacings
and the second one invariant under shifts by one lattice spacing. The action
(\ref{act1}) can be cast into a form where both the $U(1)$ and $Z_{2}$
symmetries are manifest, through the Villain approximation \cite{villain2}.
In this way the spin-wave and the vortex contributions can be separated.
Furthermore, by integrating out the spin waves, the resulting vortex
contribution can be rewritten as a fractionally charged Coulomb gas (CG)
defined on the dual lattice \cite{fradkin}:
\begin{equation}
H=-J\sum_{r,r^{\prime }}\left( m\left( r\right) +f\right) G\left(
r,r^{\prime }\right) \left( m\left( r^{\prime }\right) +f\right) .
\label{act2}
\end{equation}
Here we have $\lim_{\left| r-r^{\prime }\right| \rightarrow \infty }G\left(
r,r^{\prime }\right) =\log \left| r-r^{\prime }\right| +\frac{1}{2}\pi $ and
the neutrality condition $\sum_{r}\left( m\left( r\right) +f\right)
=\sum_{r}n\left( r\right) =0$ must be satisfied. It is now evident that the
ground state for $f=\frac{1}{2}$ consists of an alternating lattice of
logarithmically interacting $\pm \frac{1}{2}$ charges and is doubly
degenerate. Such a model exhibits two possible phase transitions, an Ising
and a vortex-unbinding one \cite{ktou}, and their relative order has been
deeply studied in the literature \cite{ffxy1}.

The $FFXY$ model has been studied analyzing other models which have the same
$U(1)\otimes Z_{2}$ degenerate ground state and are believed to be in the
same universality class. In particular it can be reformulated in terms of a
system of two coupled $XY$ models with a symmetry breaking term \cite
{granato1}:
\begin{equation}
H=A\left[ \sum_{i=1,2}\sum_{\left\langle r,r^{^{\prime }}\right\rangle }\cos
\left( \varphi ^{(i)}(r)-\varphi ^{(i)}(r^{^{\prime }})\right) \right]
+h\sum_{r}\cos 2\left( \varphi ^{(1)}(r)-\varphi ^{(2)}(r)\right) .
\label{H-GK}
\end{equation}
The limit $h\rightarrow 0$ corresponds to a full decoupling of the fields $%
\varphi ^{(i)}$, $i=1,2$, so giving rise to a CFT with central charge $c=2$
which describes two independent classical $XY$ models (in the continuum
limit). In the $h\rightarrow \infty $ limit the two phases $\varphi ^{(i)}$,
$i=1,2$ are locked \cite{granato1}, i. e. $\varphi ^{(1)}(r)-\varphi
^{(2)}(r)=\pi j$, $j=1,2$; as a consequence the model gains a symmetry $%
U(1)\otimes Z_{2}$ and its Hamiltonian renormalizes towards a model
described by:
\begin{equation}
H=H\left( h\rightarrow \infty \right) =A\sum_{\left\langle r,r^{^{\prime
}}\right\rangle }\left( 1+s_{r}s_{r^{^{\prime }}}\right) \cos \left( \varphi
^{(1)}(r)-\varphi ^{(1)}(r^{^{\prime }})\right)
\end{equation}
where $\varphi ^{(1)}(r)$ and $s_{r}=\cos \pi j=\pm 1$ are planar and Ising
spins respectively. In this way a model is obtained which is consistent with
the required symmetry, the $XY$-Ising one, and whose Hamiltonian has the
general form \cite{granato}:
\begin{equation}
H_{XY-I}=\sum_{\left\langle r,r^{^{\prime }}\right\rangle }\left[ A\left(
1+s_{r}s_{r^{^{\prime }}}\right) \cos \left( \varphi ^{(1)}(r)-\varphi
^{(1)}(r^{^{\prime }})\right) +Cs_{r}s_{r^{^{\prime }}}\right] .
\label{hamil4}
\end{equation}
In the presence of such a symmetry we can use the $m$-reduction technique
\cite{cgm4} which has been successfully applied to a quantum Hall fluid in
\cite{cgm2,cgm4,noi1,noi2} and to a fully frustrated Josephson junction
ladder with Mobius boundary conditions in \cite{noi3,noi4,noi6}.

Let us now have a look at the full spectrum of excitations of the $FFXY$
model: vortices, domain walls, kinks and antikinks. Vortices are point-like
defects such that the phase rotates by $\pm 2\pi $ in going around them \cite
{ktou}. A domain wall is a topological excitation of the double degenerate
ground state and it can be defined as a line of links, each one separating
two plaquettes with the same chirality. So through a domain wall the
alternating structure (the checkerboard pattern) of the ground state is
lost. Kinks and antikinks are excitations which live on the domain walls and
are described by fractional vortices with $+1/2$ and $-1/2$ topological
charge. As we will see, domain wall excitations can be generated from the
ground state by closing the lattice. This is our strategy: the JJA under
study can be closed and arranged in a Corbino disk geometry, which is the
relevant geometry for the implementation of a protected qubit. Imposing the
coincidence between the internal and the external edge we obtain the
discretized analogue of a torus. In this way we generate a variety of
boundary conditions which we can classify by noticing that, for the ground
state, topologically inequivalent circumstances arise for even or odd number
of plaquettes along the cycles of the torus. In the even case the end
plaquettes on the opposite sides of the lattice have opposite chirality,
while in the odd case they have the same chirality. So the ground state on a
lattice maps into the ground state on the torus only if the torus has an
even number of plaquettes along the two cycles. On the other hand a straight
domain wall is generated along any cycle of the torus which corresponds to
opposite sides of the lattice separated by an odd number of plaquettes; all
the possible cases are illustrated in Figs. 1,...,4 in the so-called {\it %
minimal configurations}. Such a behavior has to be taken into account by
opportune boundary conditions on the fields $\varphi ^{(i)}$, $i=1,2$ at the
edges of the finite lattice. These non trivial boundary conditions naturally
arise when we implement the $m$-reduction procedure in the discrete case. So
the closed geometry gives rise to non-trivial topological properties, which
appear deeply related to the twofold degeneracy of the ground state and are
the source of protection from external perturbations.

The next step is to show how our $m$-reduction procedure will give rise in a
natural way to such non trivial topological properties for the closed JJA.
The first step will be to implement the $m$-reduction in the discrete case
and then to perform the continuum limit first on the plane and then on the
torus. In this way, by using the powerful techniques of the CFT, we will be
able to propose a qubit device protected from decoherence.

\section{The $m$-reduction procedure for the JJA: a summary}

Here we briefly summarize the main results of our theory, the TM,
for the fully frustrated JJA \cite{noi}. We first construct the
bosonic theory on the plane and show that it gives rise to a
system with the required $U\left( 1\right) \otimes Z_{2}$
symmetry. That allows us to describe the JJA excitations in terms
of the primary fields $V_{\alpha }\left( z\right) $. Then we
implement a discrete version of the $m$-reduction procedure in
order to clarify the central role played by the closed geometry
and by the non-trivial boundary conditions in the description of
the excitations spectrum. The topological structure of the vacua
on the torus will be outlined in Section 4 together with the
corresponding physical configurations of the JJA, the detailed
expressions of the conformal blocks being reported in the
Appendix.

Let us focus on the $m$-reduction procedure \cite{cgm4} for the special $m=2$
case (see Ref. \cite{cgm2} for the general case), since we are interested in
a system with $U\left( 1\right) \otimes Z_{2}$ symmetry and choose the
``bosonic''\ theory \cite{noi}, which well adapts to the description of a
system with Cooper pairs of electric charge $2e$ in the presence of
non-trivial boundary conditions \cite{noi1}, i.e. a fully frustrated JJA. As
a result of the $2$-reduction procedure \cite{cgm2}\cite{cgm4} we get a $c=2$
orbifold CFT, the TM, whose fields have well defined transformation
properties under the discrete $Z_{2}$ (twist) group, which is a symmetry of
the TM. Its primary fields content can be expressed in terms of a $Z_{2}$%
-invariant scalar field $X(z)$, given by
\begin{equation}
X(z)=\frac{1}{2}\left( Q^{(1)}(z)+Q^{(2)}(z)\right) ,  \label{X}
\end{equation}
describing the continuous phase sector of the theory, and a twisted field
\begin{equation}
\phi (z)=\frac{1}{2}\left( Q^{(1)}(z)-Q^{(2)}(z)\right) ,  \label{phi}
\end{equation}
which satisfies the twisted boundary conditions $\phi (e^{i\pi }z)=-\phi (z)$
\cite{cgm2}. The whole TM theory decomposes into a tensor product of two
CFTs, a twisted invariant one with $c=\frac{3}{2}$ and the remaining $c=%
\frac{1}{2}$ one realized by a Majorana fermion in the twisted sector. In
the $c=\frac{3}{2}$ sub-theory the primary fields are composite vertex
operators $V\left( z\right) =U_{X}^{\alpha _{l}}\left( z\right) \psi \left(
z\right) $ or $V_{qh}\left( z\right) =U_{X}^{\alpha _{l}}\left( z\right)
\sigma \left( z\right) $, where
\begin{equation}
U_{X}^{\alpha _{l}}\left( z\right) =\frac{1}{\sqrt{z}}:e^{i\alpha _{l}X(z)}:
\label{char}
\end{equation}
is the vertex of the continuous\ sector with $\alpha _{l}=\frac{l}{2}$, $%
l=1,...,4$ for the $SU(2)$ Cooper pairing symmetry used here. Regarding the
other\ component, the highest weight state in the isospin sector, it can be
classified by the two chiral operators:
\begin{equation}
\psi \left( z\right) =\frac{1}{2\sqrt{z}}\left( :e^{i\sqrt{2}\phi \left(
z\right) }:+:e^{i\sqrt{2}\phi \left( -z\right) }:\right) ,~~~~~~\overline{%
\psi }\left( z\right) =\frac{1}{2\sqrt{z}}\left( :e^{i\sqrt{2}\phi \left(
z\right) }:-:e^{i\sqrt{2}\phi \left( -z\right) }:\right) ;  \label{neu1}
\end{equation}
which correspond to two $c=\frac{1}{2}$ Majorana fermions with Ramond
(invariant under the $Z_{2}$ twist) or Neveu-Schwartz ($Z_{2}$ twisted)
boundary conditions \cite{cgm2}\cite{cgm4} in a fermionized version of the
theory. The Ramond fields are the degrees of freedom which survive after the
tunnelling and the parity symmetry, which exchanges the two Ising fermions,
is broken. Besides the fields appearing in eq. (\ref{neu1}), there are the $%
\sigma \left( z\right) $ fields, also called the twist fields, which appear
in the quasi-hole primary fields $V_{qh}\left( z\right) $. The twist fields
have non local properties and decide also for the non trivial properties of
the vacuum state, which in fact can be twisted or not in our formalism.
Indeed the whole TM theory decomposes into a tensor product of two CFTs, a
twisted invariant one with $c=3/2$ (the Moore-Read (MR) theory with symmetry
$U(1)\otimes Z_{2}$) and the remaining $c=1/2$ one realized by a Majorana
fermion in the twisted sector. Such a factorization can be unambiguously
pointed out on the torus topology as we will show in the following \cite
{cgm4}. But now let us focus on a discrete version of the procedure just
outlined in order to clarify the role of the closed geometry and of the
non-trivial boundary conditions in the description of the full spectrum of
excitations of the model \cite{noi}. To such an extent let $(-L/2,0)$, $%
(L/2,0)$, $(L/2,L)$, $(-L/2,L)$ be the corners of the square lattice ${\cal L%
}$ and assume that the fields $\varphi ^{(i)}$, $i=1,2$ satisfy the
following boundary conditions:
\begin{equation}
\varphi ^{(1)}(r)=\varphi ^{(2)}(r)\text{\ \ \ for }r\in {\cal L}\cap {\bf x}%
,
\end{equation}
where ${\bf x}$ is the $x$ axis. The above boundary conditions allow us to
consider the two fields $\varphi ^{(1)}$ and $\varphi ^{(2)}$ on the square
lattice ${\cal L}$ as the folding of a single field ${\cal Q}$, defined on
the lattice ${\cal L}_{0}$ with corners $(-L/2,-L),$ $(L/2,-L)$, $(L/2,L)$, $%
(-L/2,L)$. More precisely we define the field ${\cal Q}$ as:
\begin{equation}
{\cal Q}(r)=\left\{
\begin{array}{c}
\varphi ^{(1)}(r)\text{ \ \ \ \ \ \ \ \ \ \ for }r\in {\cal L}\cap {\cal L}%
_{0}, \\
\varphi ^{(2)}(-r)\text{ \ \ \ for }r\in (-{\cal L)}\cap {\cal L}_{0}.
\end{array}
\right.
\end{equation}
We can implement now a discrete version of the $m$-reduction procedure ($m=2$%
) by defining the fields:
\begin{equation}
{\cal X}(r)=\frac{1}{2}\left( {\cal Q}(r)+{\cal Q}(-r)\right) ,
\label{X-scalar}
\end{equation}
\begin{equation}
\Phi (r)=\frac{1}{2}\left( {\cal Q}(r)-{\cal Q}(-r)\right) ,
\label{ph-scalar}
\end{equation}
where $r\in {\cal L}_{0}$. The resemblance with the continuum version of the
procedure is evident and the fields ${\cal X}$ and $\Phi $ are symmetric and
antisymmetric with respect to the action of the generator $g:r\rightarrow -r$
of the discrete group $Z_{2}$. The Hamiltonian in Eq. (\ref{H-GK}) can be
rewritten in terms of these fields and, for $h=0$, it becomes:
\begin{equation}
H=2A\sum_{\left\langle r,r^{^{\prime }}\right\rangle \in {\cal L}}\cos
\left( {\cal X}(r)-{\cal X}(r^{^{\prime }})\right) \cos \left( \Phi (r)-\Phi
(r^{^{\prime }})\right) ,  \label{dham2}
\end{equation}
which in the continuum limit corresponds to the action of our TM model:
\begin{equation}
{\cal A}=\int \left[ \frac{1}{2}\left( \partial {\cal X}\right) ^{2}+\frac{1%
}{2}\left( \partial \Phi \right) ^{2}\right] d^{2}x\text{.}
\end{equation}
It is worth pointing out that the fields ${\cal X}$ and $\Phi $ are scalar
fields and so the chiral fields defined by Eqs. (\ref{X}), (\ref{phi}) can
be seen as their chiral components. Moreover the group $Z_{2}$ is a discrete
symmetry group, indeed both $H$ and ${\cal A}$ are invariant under its
action.

\section{JJA on the torus topology: magnetic translations and wave functions}

The non perturbative ground state wave functions of the JJA system in the
torus topology represent coherent states of Cooper pairs on the torus. It
can be inferred that such ground states can be expressed as correlation
functions of the primary fields describing the elementary particles (Cooper
pairs) of JJA. In the following we will explicitly show that the characters
of the theory are in one to one correspondence with such ground states. In
particular they describe the components of the ``center of charge'' for the
corresponding ground state wave functions \cite{gerardo}, encoding all the
topological properties of the JJA system. To such an extent let us define
for a single Cooper pair on a torus $a\times b$ an effective mean-field
Hamiltonian of the kind $H\left( x,y\right) =H_{0}\left( x,y\right) +V\left(
x,y\right) $, where $H_{0}\left( x,y\right) =\left[ -i\hbar \overrightarrow{%
\nabla }-2e\overrightarrow{A}/c\right] ^{2}/2m$ is the Hamiltonian in the
presence of an uniform magnetic field and $V\left( x,y\right) $ is a
mean-field scalar potential such that $V\left( x,y\right) =V\left(
x+a,y\right) =V\left( x,y+b\right) $. It is now possible to define the
magnetic translations operators $\widetilde{{\cal S}}=e^{i\theta _{x}a/\hbar
}$ and $\widetilde{{\cal T}}=e^{i\theta _{y}b/\hbar }$ along the two cycles $%
A$ (i. e. the real axis ${\bf x}$) and $B$ (i. e. the imaginary axis ${\bf y}
$) of the torus respectively, where:
\begin{equation}
\begin{array}{cc}
\theta _{x}=\pi _{x}-\frac{2e}{c}By=-i\hbar \partial _{x}, & \theta _{y}=\pi
_{y}+\frac{2e}{c}Bx=-i\hbar \partial _{y}+\frac{2e}{c}Bx
\end{array}
\label{mgo1}
\end{equation}
and the gauge choice $\overrightarrow{A}\left( x,y\right) =\left(
-By,0\right) $ has been made. They satisfy the relations:
\begin{equation}
\begin{array}{cc}
\left[ \widetilde{{\cal S}},{\cal H}\left( x,y\right) \right] =\left[
\widetilde{{\cal T}},{\cal H}\left( x,y\right) \right] =0, & \widetilde{%
{\cal S}}\widetilde{{\cal T}}=e^{2\pi i\Phi _{ab}/\Phi _{0}}\widetilde{{\cal %
T}}\widetilde{{\cal S}}
\end{array}
,  \label{mgo2}
\end{equation}
where $\Phi _{ab}$\ is the magnetic flux threading the torus surface, and
their action on the wave functions can be defined as:
\begin{equation}
\begin{array}{cc}
{\bf \ }\widetilde{{\cal S}}\varphi \left( x,y\right) =\varphi \left(
x+a,y\right) , & \widetilde{{\cal T}}\varphi \left( x,y\right) =e^{2\pi
iBbx/\Phi _{0}}\varphi \left( x,y+b\right)
\end{array}
.  \label{mgo3}
\end{equation}
Now for $\Phi _{ab}=M\Phi _{0}$ (i.e. when the magnetic flux $\Phi _{ab}$\
is an integer number $M$ of flux quanta $\Phi _{0}=\frac{hc}{2e}$) the
condition $\left[ \widetilde{{\cal S}},\widetilde{{\cal T}}\right] =0$ holds
and we can simultaneously diagonalize the operators $H\left( x,y\right) $, $%
\widetilde{{\cal S}}$, $\widetilde{{\cal T}}$.\ By introducing adimensional
coordinates on the torus, Eqs. (\ref{mgo3}) can be rewritten as:
\begin{equation}
\begin{array}{cc}
{\bf \ }\widetilde{{\cal S}}\varphi \left( \omega \right) =\varphi \left(
\omega +1\right) , & \widetilde{{\cal T}}\varphi \left( \omega \right)
=e^{2\pi iMx}\varphi \left( \omega +\tau \right)
\end{array}
,  \label{mgo4}
\end{equation}
where $\omega =x+\tau y$, $x\in \left[ 0,1\right] $, $y\in \left[ 0,1\right]
$. One can look for eigenfunctions $\varphi \left( \omega \right) =e^{i\pi
My^{2}\tau }f\left( \omega \right) $ and define magnetic translation
operators ${\cal S}_{\alpha },{\cal T}_{\alpha }$\ acting only on the
corresponding holomorphic part $f\left( \omega \right) $:
\begin{equation}
\begin{array}{cc}
{\bf \ }{\cal S}_{\alpha }f\left( \omega \right) =f\left( \omega +\alpha
\right) , & {\cal T}_{\alpha }f\left( \omega \right) =e^{i\pi M\left( \alpha
^{2}\tau +2\alpha \omega \right) }f\left( \omega +\alpha \tau \right)
\end{array}
.  \label{mgo5}
\end{equation}
In this way Eqs. (\ref{mgo4}) become:
\begin{equation}
\begin{array}{cc}
{\bf \ }\widetilde{{\cal S}}\varphi \left( \omega \right) =e^{i\pi
My^{2}\tau }{\cal S}_{1}f\left( \omega \right) , & \widetilde{{\cal T}}%
\varphi \left( \omega \right) =e^{i\pi My^{2}\tau }{\cal T}_{1}f\left(
\omega \right)
\end{array}
.  \label{mgo6}
\end{equation}
It is straightforward to show that the ground state wave function for $M$
Cooper pairs is a coherent state given by:
\begin{equation}
\psi _{a}\left( \omega _{1},...,\omega _{M}\right) =e^{i\pi M\tau
\sum_{i=1}^{M}y_{i}^{2}}f_{a}\left( \omega _{1},...,\omega _{M}\right) ,
\label{mgo7}
\end{equation}
\begin{equation}
f_{a}\left( \omega _{1},...,\omega _{M}\right) =\prod_{i<j=1}^{M}\left[
\frac{\theta _{1}\left( \omega _{ij},\tau \right) }{\theta _{1}^{^{\prime
}}\left( 0,\tau \right) }\right] ^{4}\chi _{a}\left( \omega |\tau \right) ,
\label{mgo8}
\end{equation}
where $\omega =\sum_{i=1}^{M}\omega _{i}$ is the ``center of charge''
variable and the non local functions $\chi _{a}\left( \omega |\tau \right) $
are the characters of our theory (the TM), whose analytical expressions are
reported in the Appendix. In fact it can be shown that such characters are
eigenfunctions of the following generalized magnetic translations operators:
\begin{equation}
\begin{array}{cc}
{\bf \ }{\cal S}_{\alpha }=\prod_{i=1}^{M}{\cal S}_{\alpha /M}^{i}, & {\cal T%
}_{\alpha }=\prod_{i=1}^{M}{\cal T}_{\alpha /M}^{i}
\end{array}
,  \label{mgo9}
\end{equation}
where $S_{\alpha /M}^{i}$ and $T_{\alpha /M}^{i}$ are the magnetic
translation operators for the single Cooper pair. In particular, by
recalling the explicit expressions of the conformal blocks (\ref{vac1})-(\ref
{vac5}) and (\ref{tw1})-(\ref{tw4}) in the four topological sectors \cite
{cgm4}, given in the Appendix, and the transformation properties of the
Theta functions \cite{theta}, we get (for the step $\alpha =1$, see also Eq.
(\ref{mgo6})):
\begin{equation}
\begin{array}{cc}
{\cal T}\tilde{\chi}_{\alpha }^{+}(w_{c}|\tau )=\tilde{\chi}_{\alpha
}^{+}(w_{c}|\tau ), & {\cal T}\tilde{\chi}_{\beta }^{+}(w_{c}|\tau )=\tilde{%
\chi}_{\beta }^{+}(w_{c}|\tau ), \\
{\cal T}\tilde{\chi}_{\gamma }^{+}(w_{c}|\tau )=\tilde{\chi}_{\gamma
}^{+}(w_{c}|\tau ), &
\end{array}
\end{equation}
for the $P-P$ sector,
\begin{equation}
\begin{array}{cc}
{\cal T}\tilde{\chi}_{\left( +\right) }^{-}(w_{c}|\tau )=i\tilde{\chi}%
_{\left( +\right) }^{-}(w_{c}|\tau ), & {\cal T}\tilde{\chi}_{\left(
-\right) }^{-}(w_{c}|\tau )=-i\tilde{\chi}_{\left( -\right) }^{-}(w_{c}|\tau
),
\end{array}
\end{equation}
for the $P-A$ sector, where we defined $\tilde{\chi}_{\left( \pm \right)
}^{-}(w_{c}|\tau )=\tilde{\chi}_{\left( 0\right) }^{-}(w_{c}|\tau )\pm
\tilde{\chi}_{\left( 1\right) }^{-}(w_{c}|\tau )$,
\begin{equation}
\begin{array}{cc}
{\cal T}\chi _{(0)}^{+}(w_{c}|\tau )=\chi _{(0)}^{+}(w_{c}|\tau ), & {\cal T}%
\chi _{(1)}^{+}(w_{c}|\tau )=\chi _{(1)}^{+}(w_{c}|\tau ),
\end{array}
\end{equation}
for the $A-P$ sector, and
\begin{equation}
\begin{array}{cc}
{\cal T}\chi _{(0)}^{-}(w_{c}|\tau )=i\chi _{(0)}^{-}(w_{c}|\tau ), & {\cal T%
}\chi _{(1)}^{-}(w_{c}|\tau )=i\chi _{(1)}^{-}(w_{c}|\tau ),
\end{array}
\end{equation}
for the $A-A$ sector respectively. From such a picture it is evident then
how the degeneracy of the non perturbative ground states is closely related
to the number of primary states and that our characters represent highly non
local functions: all the topological properties of our system are codified
in such functions.

Let us now focus on the general topological structure of the vacua on the
torus and on their corresponding physical JJA realization in relation to the
various boundary conditions imposed by closing the square lattice. Let $A$
be the cycle of the torus which surrounds the hole of the Corbino disk and $%
B $ the cycle in the radial direction. On a pure topological ground we
expect for the torus a doubling of the ground state degeneracy, which can be
seen at the level of the conformal blocks (characters) of our TM (see
Appendix). Indeed we get for periodic boundary conditions (i. e. an even
number of plaquettes) along the $A$-cycle an untwisted sector, $P-P$\ and $%
P-A$, characterized by periodic ($P$) and twisted ($A$) boundary conditions
along the $B$-cycle respectively, described by the four conformal blocks (%
\ref{vac1})-(\ref{vac4}); likewise we get for twisted boundary conditions
(i. e. an odd number of plaquettes) along the $A$-cycle a twisted sector, $%
A-P$\ and $A-A$, characterized by periodic and twisted boundary conditions
along the $B$-cycle respectively, described by the four conformal blocks (%
\ref{tw1})-(\ref{tw4}). The four different topological sectors just obtained
can be put in correspondence with four minimal configurations for the closed
fully frustrated JJA as follows.

Regarding the untwisted sector, in Fig. 1 we show the two vacua
corresponding to the $P-P$ sector, characterized by a JJA with $N=4$
plaquettes along the circular direction and $N=2$ plaquettes along the
radial direction, $4$ and $2$ being the minimum number of plaquettes
respectively needed in order to fulfill the even-even boundary conditions.
Then in Fig. 2 the two vacua in the $P-A$ sector, implemented by a JJA with
even-odd boundary conditions along the two cycles respectively, are
sketched.
\begin{figure}[tbp]
\includegraphics[height=.3\textheight]{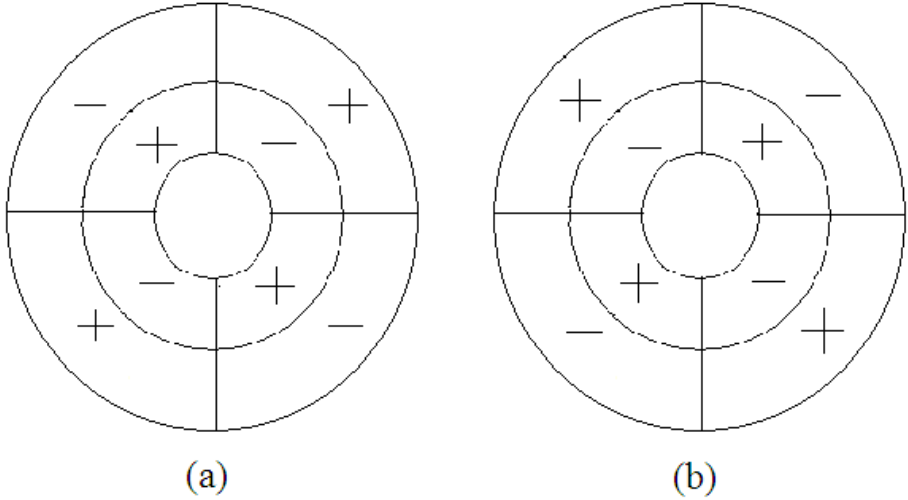}
\caption{The two vacua in the $P-P$ sector.}
\end{figure}
\begin{figure}[tbp]
\includegraphics[height=.3\textheight]{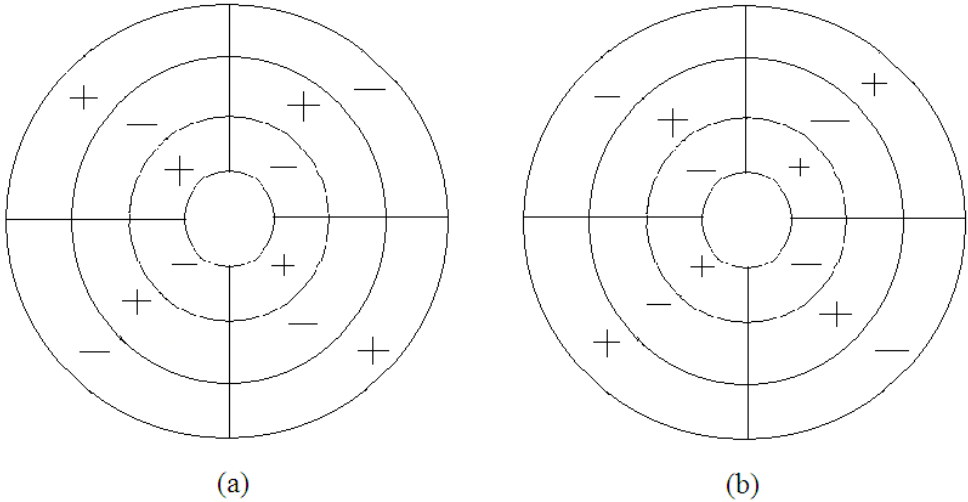}
\caption{The two vacua in the $P-A$ sector. A straight domain wall is
generated along the $A-$cycle.}
\end{figure}

Regarding the twisted sector, we get for the $A-P$ and the $A-A$ sectors the
configurations shown in Figs. 3 and 4 respectively, characterized by the
required odd-even and odd-odd boundary conditions.
\begin{figure}[tbp]
\includegraphics[height=.3\textheight]{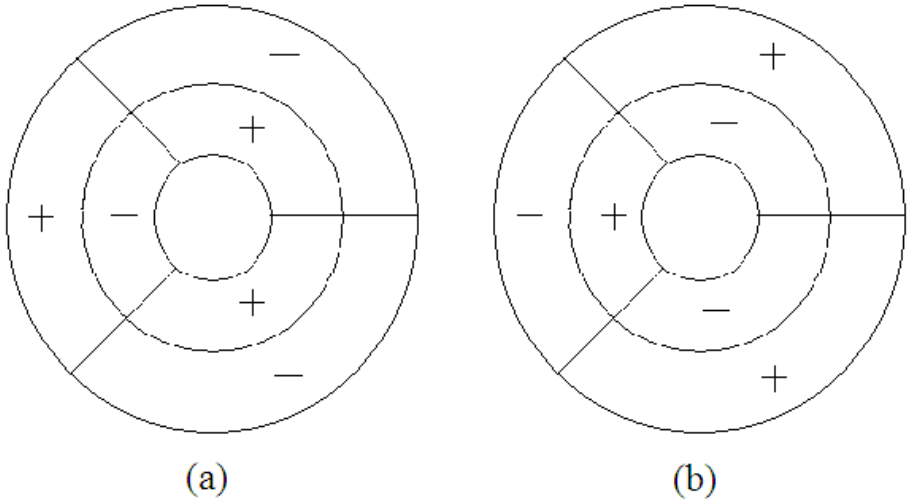}
\caption{The two vacua in the $A-P$ sector. A straight domain wall is
generated along the $B-$cycle.}
\end{figure}
\begin{figure}[tbp]
\includegraphics[height=.3\textheight]{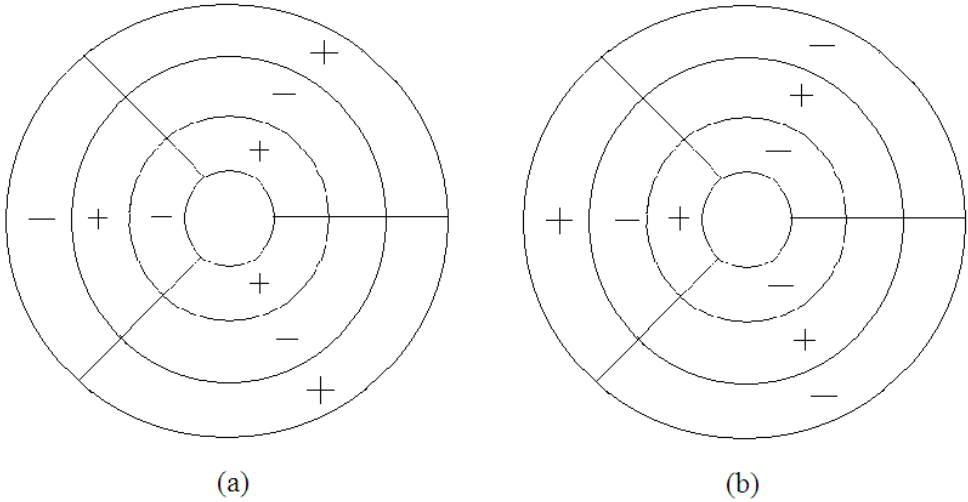}
\caption{The two vacua in the $A-A$ sector. A straight domain wall is
generated along \ both the $A-$cycle and the $B-$cycle.}
\label{figure1}
\end{figure}

In the next Section the relevant two states of the qubit will be found and
will lead us to propose the configuration shown in Fig. 3 as the minimal one
required to build a reliable qubit device protected from decoherence.

\section{Stability and transformation properties of the ground states: the
physical qubit}

In this Section we will study the stability and transformation properties of
the four ground states of the JJA arranged in the Corbino disk geometry
under an adiabatic elementary flux change ($\pm \frac{hc}{2e}$) through the
central hole. Because of the finite energy gap to fractionally charged
excitation states (in complete analogy with the presence of a gap separating
the ground state from higher energy states in the Laughlin Hall fluid \cite
{hall1}), such an adiabatic transformation is believed to leave the system
in a ground state which can be different from the original one, due to the
occurrence of the ground state degeneracy. This analysis will lead us to the
identification of the two states of a possible protected qubit and to the
definition of a tool for its manipulation.

We use the results provided by our TM model in Section 4 and in
the Appendix in order to analyze such properties by standard
conformal techniques. As we showed in the previous Section, in the
torus topology the characters of the theory are in one to one
correspondence with the ground states and any one of them
describes the component of the ``center of charge'' for the
corresponding ground state wave function. On a pure topological
ground the torus shows a doubling of the ground state degeneracy,
which can be seen at the level of the conformal blocks
(characters) of our TM. Indeed, as outlined in Section 4, we get \
a rich structure for the vacua corresponding to different boundary
conditions at the ends of the square lattice. That gives rise to
an untwisted sector, $P-P$\ and $P-A$, described by the four
conformal blocks (\ref{vac1})-(\ref{vac4}), and a twisted sector,
$A-P$\ and $A-A$, described by the four conformal blocks
(\ref{tw1})-(\ref{tw4}) respectively. Now we are going to extract
from such vacua the two states of the ``protected''qubit.

Let us notice that the ground state wave functions of the twisted and
untwisted sectors of the TM are characterized by different monodromy
properties along the $A$-cycle. In particular the characters of the
untwisted sector are single-valued functions along the $A$-cycle while the
characters of the twisted sector pick up a common $(-1)$ phase factor along
the $A$-cycle. Such phase factors can be interpreted as Bohm-Aharonov phases
generated while a Cooper pair is taken along the $A$-cycle. The above
observation evidences a strong difference between the two inequivalent
topological sectors, the untwisted and the twisted one respectively, on the
torus. Indeed in the twisted sector the ground state wave functions show a
non trivial behavior implying the trapping of a half flux quantum ($\frac{1}{%
2}\left( \frac{hc}{2e}\right) $) in the hole of the Corbino disk.
Instead in the untwisted sector, due to the single-valued ground
state wave functions, only integer numbers of flux quanta can be
attached to the hole. It is worth pointing out the central role
played by the {\it isospin} (or neutral) component of the TM in
producing the discussed non trivial monodromy properties. To this
end let us recall that the TM is a $c=2$ CFT, composed by a $c=1$
{\it charged} and a $c=1$ {\it isospin }CFT components, as it is
well evidenced by the characters decompositions given in Section
4. Furthermore the transport of a Cooper pair along the $A$-cycle
can be implemented by\ a simultaneous and identical translation
$\Delta
w_{c}=\Delta w_{n}=2$ of the charged\ and the isospin\ variables. The {\it %
charged} characters have trivial monodromy with respect to this
transformation, being:
\begin{equation}
K_{l}(w_{c}+2|\tau )=K_{l}(w_{c}|\tau ),\,\ \ l=0,..,3,
\end{equation}
while the {\it isospin} contribution is the one responsible for the non
trivial monodromy of the complete ground state wave functions:
\begin{equation}
\chi _{0,\frac{1}{2}}(2|\tau )=\chi _{0,\frac{1}{2}}(0|\tau )\,,\qquad \chi
_{\frac{1}{16}}(2|\tau )=(-1)\chi _{\frac{1}{16}}(0|\tau )  \label{c18}
\end{equation}
and the same is true for the characters $\bar{\chi}_{\beta }$.
Observe that the change in sign in the last relation of Eq.
(\ref{c18}) shows the presence in the spectrum of excitations
carrying fractionalized charge quanta. More precisely the presence
in the isospin component of one twist-field (with conformal
dimension $\Delta =1/16$) characterizes all the conformal blocks
of the twisted sector and accounts for the trapping of a half flux
quantum in the hole of the Corbino disk.

We are now ready to study the stability and transformation properties of the
ground state wave functions when a magnetic flux change takes place through
the central hole of the closed JJA. The above analysis shows that at the
level of the wave functions it has the effect to change the monodromy along
the $A$-cycle due to the corresponding change in the Bohm-Aharonov phase.
Such a modification can be implemented on the center of charge component of
the wave function, i.e. the characters, with a well defined transformation.
In the case of the {\it charged} component this analysis has been brought
out already in \cite{Napoli-91-92} for the quantum Hall effect. Let us adapt
here the results for the {\it charged} component of our TM. On a pure
physical ground the fact that we are considering a magnetic flux change,
which is on one side integer in the flux quantum (one flux quantum change $%
\pm \frac{hc}{2e}$) and on the other side adiabatic suggests both
that the monodromy properties do not change and that the system
remains in a degenerate ground state. Such a physical picture is
in fact confirmed for the {\it charged} component of our TM;
indeed, the flux change is
implemented on the {\it charged }characters by the transformation ${\cal T}%
_{1/2}^{c}$:
\begin{equation}
{\cal T}_{1/2}^{c}K_{l}\left( w_{c}|\tau \right) \equiv e^{\left( \frac{1}{2}%
\right) ^{2}i\pi \tau +\frac{2i\pi w_{c}}{2}}K_{l}\left( w_{c}+\frac{\tau }{2%
}|\tau \right) =K_{l+1}\left( w_{c}|\tau \right) ,\ l=0,..,3.
\end{equation}
Let us notice that the {\it charged} component wave functions realize a flip
process ($l\rightarrow l+1$) under one magnetic flux quantum change.

However the analysis for the complete TM, with {\it charged} and {\it isospin%
} components, is more involved, due to the non trivial interplay between
{\it charged} and {\it isospin} components summarized in the so-called $m$%
-ality parity rule, which characterizes the gluing condition for the {\it %
charged} and {\it isospin} excitations. The main point being the
compatibility between such parity rule and the transformation of the
complete characters of the TM under the insertion of a magnetic flux quantum
through the hole of the closed JJA, which reads as:
\begin{equation}
{\cal T}_{1/2}f(w_{n}|w_{c}|\tau )=\left. e^{2i\pi (\alpha ^{2}\tau +\alpha
(w_{n}+w_{c}))}f(w_{n}+\alpha \tau |w_{c}+\alpha \tau |\tau )\right|
_{\alpha =1/2},
\end{equation}
where $f(w_{n}=0|w_{c}|\tau )$ stays for any character of the TM.
More in detail the full list of character transformations is as
follows.

In the untwisted sector, we have that the two ground state wave functions of
the $P-A$ sector decouple, being
\begin{equation}
{\cal T}_{1/2}\tilde{\chi}_{(0)}^{-}(0|w_{c}|\tau )=0\,,\text{ \ }{\cal T}%
_{1/2}\tilde{\chi}_{(1)}^{-}(0|w_{c}|\tau )=0.  \label{t(1/2)-PA}
\end{equation}
Concerning the $P-P$ sector, we have:
\begin{equation}
{\cal T}_{1/2}\tilde{\chi}_{\alpha }^{+}(0|w_{c}|\tau )=0  \label{t(1/2)-PPa}
\end{equation}
and
\begin{equation}
{\cal T}_{1/2}\tilde{\chi}_{\beta }^{+}(0|w_{c}|\tau )=\tilde{\chi}_{\gamma
}^{+}(0|w_{c}|\tau )\,\text{ \ \ \ \ }(\ {\cal T}_{1/2}\tilde{\chi}_{\gamma
}^{+}(0|w_{c}|\tau )=\tilde{\chi}_{\beta }^{+}(0|w_{c}|\tau )\,\text{\ }).
\label{t(1/2)-PPbc}
\end{equation}
Such transformations show the instability of the $P-P$ sector
under the insertion of a flux quantum through the hole of the
closed JJA. More precisely the state $\tilde{\chi}_{\alpha
}^{+}(0|w_{c}|\tau )$ decouples while the state
$\tilde{\chi}_{\beta }^{+}(0|w_{c}|\tau )$ gets excited to the
state with a kink-antikink configuration $\tilde{\chi}_{\gamma
}^{+}(0|w_{c}|\tau )$ (see Eq. (\ref{vac5}) and comments
afterwards).

Furthermore in the twisted sector, we have that the two ground state wave
functions of the $A-A$ sector decouple, being
\begin{equation}
{\cal T}_{1/2}\chi _{(0)}^{-}(0|w_{c}|\tau )=0\,,\text{ \ }{\cal T}%
_{1/2}\chi _{(1)}^{-}(0|w_{c}|\tau )=0.  \label{t(1/2)-AA}
\end{equation}
Concerning the $A-P$ sector, we have that the two ground state wave
functions transform as:
\begin{equation}
{\cal T}_{1/2}\chi _{(0)}^{+}(0|w_{c}|\tau )=\chi _{(1)}^{+}(0|w_{c}|\tau
)\,,\text{ \ }{\cal T}_{1/2}\chi _{(1)}^{+}(0|w_{c}|\tau )=\chi
_{(0)}^{+}(0|w_{c}|\tau )\,.  \label{t(1/2)-AP}
\end{equation}
Concluding, the $A-P$ sector only is stable under the insertion of
a magnetic flux quantum through the central hole, with the two
ground states flipping one into the other under an adiabatic flux
change of $\pm \frac{hc}{2e}$. That allows us to make the
following identifications:
\begin{equation}
\left| 0\right\rangle \backsim \chi _{\left( 0\right) }^{+}(0|w_{c}|\tau
),~~~~~\left| 1\right\rangle \backsim \chi _{\left( 1\right)
}^{+}(0|w_{c}|\tau ),  \label{logical1}
\end{equation}
in terms of the ground states center of charge wave functions
(characters). Then $\left| 0\right\rangle $ and $\left|
1\right\rangle $\ are the two ground states of the closed JJA
characterized by an odd number of plaquettes along the $A$-cycle
and an even number of plaquettes along the $B$-cycle. If we choose
a closed path along the $A$-cycle, we can define a topological
invariant as the {\it size} invariant sum over all its
plaquettes $\sum_{p}\chi _{p}$, which results equal to $-1$ and $+1$ for\ $%
\left| 0\right\rangle $ and $\left| 1\right\rangle $\ respectively.

The results just obtained lead us to propose the JJA closed in a Corbino
disk geometry, characterized by an odd number of plaquettes along circular
direction and an even number of plaquettes along radial direction, as our
protected qubit. In fact the two ground states $\left| 0\right\rangle $ and $%
\left| 1\right\rangle $\ work as the two logical states of the qubit and the
required one qubit operations:
\[
\left| 0\right\rangle \rightarrow \left| 1\right\rangle ,~~~~~\left|
1\right\rangle \rightarrow \left| 0\right\rangle ,
\]
are simply implemented by insertion of a flux quantum ($\pm \frac{hc}{2e}$)
through the central hole. That provides a tool for the control and the
manipulation of our device, which is indeed a ``flux'' qubit \cite{orlando1}.

In Fig. 3 we showed the minimal configuration for such a device, that is a
closed fully frustrated JJA with $N=3$ plaquettes along the circular
direction and $N=2$ plaquettes along the radial direction, $3$ and $2$ being
the minimum odd and even number of plaquettes respectively needed in order
to fulfill all the above requests.

Given the two degenerate ground states (\ref{logical1}) and the operation
mode defined in (\ref{t(1/2)-AP}) it should be possible to prepare the qubit
in a definite state and to realize all elementary one-qubit operations. We
showed in particular how to implement the $NOT$ operation, the Hadamard gate
will be implemented in a future publication. All that is accomplished by an
adiabatic flux change of $\pm \frac{hc}{2e}$ through the central hole of the
Corbino disk.

\section{Conclusions and outlooks}

The ground states and the topological properties of fully frustrated
Josephson Junction arrays (JJA) arranged in a Corbino disk geometry for a
variety of boundary conditions have been investigated in detail, employing a
twisted CFT approach \cite{cgm2,cgm4}. In this way we built up a low-energy
effective field theory which is a topological one, due to the well known
relation with Chern-Simons gauge theories in $(2+1)$-dimensions \cite
{witten1}. The vacua obtained have been classified in four different
topological sectors, according to the different boundary conditions imposed
at the ends of the square lattice, and put in correspondence with four
minimal configurations for the closed fully frustrated JJA. Then a careful
analysis of the stability and transformation properties of the conformal
blocks corresponding to such vacua allowed us to propose a solid state
qubit, protected from decoherence, whose operation mode is based only on
adiabatic magnetic flux changes through the central hole of the Corbino disk.

Josephson junction arrays have been fabricated within the trilayer $%
Nb/Al-AlO_{x}/Nb$ or $Pb/Sn/Pb$ technology as well as all aluminum
technology and experimentally investigated \cite{exper}, but in such a case
the application of an external transverse magnetic field is needed in order
to fulfill the requirement of full frustration and that could be another
source of decoherence. It is now possible to avoid such a problem by
realizing arrays with a built-in frustration. In fact high-$T_{c}$ Josephson
junction arrays have been recently proposed \cite{rogalla}, which support
degenerate spontaneous current states in zero magnetic field due to the
presence of plaquettes containing an odd number of $\pi $-junctions \cite{pi}%
. Such unconventional junctions can be realized because of the $d$-wave
symmetry of high-$T_{c}$ superconductors \cite{kirtley}, which produces a $%
\pi $-shift in the phase of the wave function on one side of the
junction. Furthermore $\pi $-junctions can be obtained also with
superconducting-ferromagnetic-superconducting (SFS) \cite{sfs} and
superconducting-insulator-ferromagnetic-superconducting (SIFS)
\cite{sifs} structures. One can also achieve the same effect using
a barrier which effectively flips the spin of a tunneling
electron, i. e. when the barrier is made of a ferromagnetic
insulator \cite{tunnel1}, of a carbon nanotube \cite{tunnel2} or
of a quantum dot \cite{tunnel3} created by gating a semiconducting
nanowire. In this way it is possible to avoid the external
frustration bias but, in any case, external magnetic fields are
needed for control and read-out operations. So in principle an
experimental setup for the realization of our protected qubit can
be easily conceived, whose basis is a fully frustrated JJA
arranged in a Corbino disk geometry with an odd number of
plaquettes along the inner hole and an even number of them along
the radial direction.

\section*{Acknowledgments}

We warmly thank P. Minnhagen for many enlightening discussions. G. N. is
supported by the contract MEXT-CT-2006-042695.

\section*{Appendix: the TM on the torus}

In the following we summarize the whole primary field content of our theory,
the TM, on the torus topology \cite{cgm4}.

On the torus \cite{cgm4} the TM primary fields are described in terms of the
conformal blocks of the $Z_{2}$-invariant $c=\frac{3}{2}$ sub-theory and of
the non invariant $c=\frac{1}{2}$ Ising model, so reflecting the
decomposition on the plane outlined in Section 3. The characters $\bar{\chi}%
_{0}(0|\tau )$, $\bar{\chi}_{\frac{1}{2}}(0|\tau )$, $\bar{\chi}_{\frac{1}{16%
}}(0|\tau )$ express the primary fields content of the Ising model \cite{cft}
with Neveu-Schwartz ($Z_{2}$ twisted) boundary conditions \cite{cgm4}, while
\begin{eqnarray}
\chi _{(0)}^{c=3/2}(0|w_{c}|\tau ) &=&\chi _{0}(0|\tau )K_{0}(w_{c}|\tau
)+\chi _{\frac{1}{2}}(0|\tau )K_{2}(w_{c}|\tau )\,,  \label{mr1} \\
\chi _{(1)}^{c=3/2}(0|w_{c}|\tau ) &=&\chi _{\frac{1}{16}}(0|\tau )\left(
K_{1}(w_{c}|\tau )+K_{3}(w_{c}|\tau )\right) ,  \label{mr2} \\
\chi _{(2)}^{c=3/2}(0|w_{c}|\tau ) &=&\chi _{\frac{1}{2}}(0|\tau
)K_{0}(w_{c}|\tau )+\chi _{0}(0|\tau )K_{2}(w_{c}|\tau )  \label{mr3}
\end{eqnarray}
represent those of the $Z_{2}$-invariant $c=\frac{3}{2}$ CFT. They are given
in terms of a ``charged''\ $K_{\alpha }(w_{c}|\tau )$ contribution:
\begin{equation}
K_{2l+i}(w|\tau )=\frac{1}{\eta \left( \tau \right) }\;\Theta \left[
\begin{array}{c}
\frac{2l+i}{4} \\[6pt]
0
\end{array}
\right] (2w|4\tau )\,,\qquad \text{with }l=0,1\text{ and }i=0,1\,,
\label{chp}
\end{equation}
and a ``isospin''\ one $\chi _{\beta }(0|\tau )$, (the conformal blocks of
the Ising Model), where $w_{c}=\dfrac{1}{2\pi i}\,\ln z_{c}$ is the torus
variable of the ``charged''\ component while the corresponding argument of
the isospin block is $w_{n}=0$ everywhere.

If we now turn to the whole $c=2$ theory, the characters of the twisted
sector are given by:
\begin{eqnarray}
\chi _{(0)}^{+}(0|w_{c}|\tau ) &=&\bar{\chi}_{\frac{1}{16}}(0|\tau )\left(
\chi _{0}+\chi _{\frac{1}{2}}\right) (0|\tau )\left( K_{0}+K_{2}\right)
(w_{c}|\tau ),  \label{tw1} \\
\chi _{(1)}^{+}(0|w_{c}|\tau ) &=&\chi _{\frac{1}{16}}(0|\tau )\left( \bar{%
\chi}_{0}+\bar{\chi}_{\frac{1}{2}}\right) (0|\tau )\left( K_{1}+K_{3}\right)
(w_{c}|\tau ),  \label{tw2}
\end{eqnarray}
for the $A-P$ sector and by:
\begin{eqnarray}
\chi _{(0)}^{-}(0|w_{c}|\tau ) &=&\bar{\chi}_{\frac{1}{16}}(0|\tau )\left(
\chi _{0}-\chi _{\frac{1}{2}}\right) (0|\tau )\left( K_{0}-K_{2}\right)
(w_{c}|\tau ),  \label{tw3} \\
\chi _{(1)}^{-}(0|w_{c}|\tau ) &=&\chi _{\frac{1}{16}}(0|\tau )\left( \bar{%
\chi}_{0}-\bar{\chi}_{\frac{1}{2}}\right) (0|\tau )\left( K_{1}+K_{3}\right)
(w_{c}|\tau ),  \label{tw4}
\end{eqnarray}
for the $A-A$ one. Furthermore the characters of the untwisted sector are
\cite{cgm4}:
\begin{align}
\tilde{\chi}_{(0)}^{-}(0|w_{c}|\tau )& =\left( \bar{\chi}_{0}\chi _{0}-\bar{%
\chi}_{\frac{1}{2}}\chi _{\frac{1}{2}}\right) (0|\tau )K_{0}(w_{c}|\tau
)+\left( \bar{\chi}_{0}\chi _{\frac{1}{2}}-\bar{\chi}_{\frac{1}{2}}\chi
_{0}\right) (0|\tau )K_{2}\,(w_{c}|\tau ),  \label{vac1} \\
\tilde{\chi}_{(1)}^{-}(0|w_{c}|\tau )& =\left( \bar{\chi}_{0}\chi _{\frac{1}{%
2}}-\bar{\chi}_{\frac{1}{2}}\chi _{0}\right) (0|\tau )K_{0}(w_{c}|\tau
)+\left( \bar{\chi}_{0}\chi _{0}-\bar{\chi}_{\frac{1}{2}}\chi _{\frac{1}{2}%
}\right) (0|\tau )K_{2}\,(w_{c}|\tau ),
\end{align}
for the $P-A$ sector while for the $P-P$ sector we have:
\begin{align}
\tilde{\chi}_{\alpha }^{+}(0|w_{c}|\tau )& =\frac{1}{2}\left( \bar{\chi}_{0}-%
\bar{\chi}_{\frac{1}{2}}\right) (0|\tau )\left( \chi _{0}-\chi _{\frac{1}{2}%
}\right) (0|\tau )(K_{0}-K_{2})(w_{c}|\tau )\,, \\
\tilde{\chi}_{\beta }^{+}(0|w_{c}|\tau )& =\frac{1}{2}\left( \bar{\chi}_{0}+%
\bar{\chi}_{\frac{1}{2}}\right) (0|\tau )\left( \chi _{0}+\chi _{\frac{1}{2}%
}\right) (0|\tau )(K_{0}+K_{2})(w_{c}|\tau ),  \label{vac4}
\end{align}
and
\begin{equation}
\tilde{\chi}_{\gamma }^{+}(0|w_{c}|\tau )=\bar{\chi}_{\frac{1}{16}}(0|\tau
)\chi _{\frac{1}{16}}(0|\tau )\left( K_{1}+K_{3}\right) (w_{c}|\tau ).
\label{vac5}
\end{equation}
Let us point out that the above factorization expresses the parity selection
rule ($m$-ality), which gives a gluing condition for the ``charged''\ and
``isospin''\ excitations.

Furthermore, in the $P-P$ sector, unlike for the other sectors, modular
invariance constraint requires the presence of three different characters.
The isospin operator content of the character $\tilde{\chi}_{\gamma
}^{+}(0|w_{c}|\tau )$ clearly evidences its peculiarity with respect to the
other states of the periodic (even ladder) case. Indeed it is characterized
by two twist fields ($\Delta =1/16$) in the isospin components. The
occurrence of the double twist in the state described by $\tilde{\chi}%
_{\gamma }^{+}(0|w_{c}|\tau )$ is simply the reason why such a
state is a periodic state. Indeed, being an isospin twist field
the representation in the continuum limit of a kink (or
equivalently a half flux quantum trapping), the double twist
corresponds to a double half flux quantum trapping, i.e. one flux
quantum, typical of the periodic configuration. Indeed the state
described by $\tilde{\chi}_{\gamma }^{+}(0|w_{c}|\tau )$ embeds in
the continuum limit a kink-antikink excitation, i.e. it represents
an excited state in the $P-P$ sector. In this way, as it happens
for all the other sectors, the $P-P$ sector is left with just two
degenerate ground states ($\tilde{\chi}_{\alpha }^{+}(0|w_{c}|\tau
)$ and $\tilde{\chi}_{\beta }^{+}(0|w_{c}|\tau )$) and, as
expected on a pure topological base, the ground state degeneracy
in the torus topology is the double of that of the disk.

Finally the partition function of the TM model on the torus has
the following factorized form (see \cite{cgm4}):
\begin{equation}
Z(w_{c}|\tau )=Z^{MR}(w_{c}|\tau )Z_{\overline{I}}(\tau )\text{,}
\end{equation}
in terms of the Moore-Read partition function $Z^{MR}$ ($c=3/2$) and of the
Ising partition function $Z_{\overline{I}}$ ($c=1/2$). They have the
following expression:
\begin{equation}
Z^{MR}(w_{c}|\tau )=\left| \chi _{0}^{MC}(0|w_{c}|\tau )\right| ^{2}+\left|
\chi _{1}^{MC}(0|w_{c}|\tau )\right| ^{2}+\left| \chi _{2}^{MC}(0|w_{c}|\tau
)\right| ^{2},
\end{equation}
\begin{equation}
Z_{\overline{I}}(\tau )=\left| \bar{\chi}_{0}(\tau )\right| ^{2}+\left| \bar{%
\chi}_{\frac{1}{2}}(\tau )\right| ^{2}+\left| \bar{\chi}_{\frac{1}{16}}(\tau
)\right| ^{2}.
\end{equation}


\begin{thebibliography}{99}
\bibitem{noi6}  G. Cristofano, V. Marotta, A. Naddeo, G. Niccoli, {\it Phys.
Lett. A }(2008) in print.

\bibitem{cgm2}  G. Cristofano, G. Maiella, V. Marotta, {\it Mod. Phys. Lett.
A}{\bf \ 15} (2000) 1679.

\bibitem{cgm4}  G. Cristofano, G. Maiella, V. Marotta, G. Niccoli, {\it %
Nucl. Phys. B}{\bf \ 641 }(2002) 547.

\bibitem{noi4}  G. Cristofano, V. Marotta, A. Naddeo, G. Niccoli, {\it Eur.
Phys. J. B}{\bf \ 49 }(2006) 83.

\bibitem{noi3}  G. Cristofano, V. Marotta, A. Naddeo, {\it J. Stat. Mech.:
Theor. Exper. }(2005) P03006.

\bibitem{revqubit}  Y. Makhlin, G. Schon, A. Shnirman, {\it Rev. Mod. Phys. }%
{\bf 73 }(2001) 357; M. H. Devoret, A. Wallraff, J. M. Martinis, {\it %
cond-mat}/0411174; G. Wendin, V. S. Shumeiko, in Handbook of Theoretical and
Computational Nanotechnology, M. Rieth and W. Schommers (Eds.), American
Scientific Publishers, Los Angeles, 2006, Vol. 3, p.223.

\bibitem{shor}  P. W. Shor, {\it Phys. Rev. A }{\bf 52} (1995) 2493.

\bibitem{kitaev}  A. Kitaev, {\it Ann. Phys. }{\bf 303 }(2003) 2.

\bibitem{wen}  X. G. Wen, Q. Niu, {\it Phys. Rev. B }{\bf 41 }(1990) 9377.

\bibitem{wen1}  X. G. Wen, {\it Phys. Lett. A }{\bf 300 }(2002) 175; X. G.
Wen, {\it Phys. Rev. B }{\bf 65 }(2002) 165113; T. Senthil, M. P. A. Fisher,
{\it Phys. Rev. B }{\bf 62 }(2000) 7850.

\bibitem{nussinov}  Z. Nussinov, G. Ortiz, {\it cond-mat}/0702377.

\bibitem{fisher1}  T. Senthil, M. P. A. Fisher, {\it Phys. Rev. Lett. }{\bf %
86} (2001) 292; T. Senthil, M. P. A. Fisher, {\it Phys. Rev. B }{\bf 63}
(2001) 134521; T. H. Hansson, V. Oganesyan, S. L. Sondhi, {\it Ann. Phys. }%
{\bf 313} (2004) 497.

\bibitem{pasquale}  M. C. Diamantini, P. Sodano, C. A. Trugenberger, {\it %
Nucl. Phys. B }{\bf 474 }(1996) 641; M. C. Diamantini, P. Sodano, C. A.
Trugenberger, {\it J. Phys. A }{\bf 39 }(2006) 253; M. C. Diamantini, P.
Sodano, C. A. Trugenberger, {\it Eur. Phys. J. B }{\bf 53 }(2006) 19.

\bibitem{nonabelian}  S. Das Sarma, M. Freedman, C. Nayak, S. H. Simon, A.
Stern, {\it cond-mat}/0707.1889; S. Bravyi, {\it Phys. Rev. A }{\bf 73}
(2006) 042313; G. K. Brennen, J. K. Pachos, {\it quant-ph}/0704.2241; S. Das
Sarma, M. Freedman, C. Nayak, {\it Phys. Rev. Lett. }{\bf 94} (2005) 166802;
M. Freedman, C. Nayak, K. Walker, {\it Phys. Rev. B }{\bf 73} (2006) 245307;
L. Georgiev, {\it Phys. Rev. B }{\bf 74} (2006) 235112.

\bibitem{metaloxide}  N. Read, D. Green, {\it Phys. Rev. B }{\bf 61} (2000)
10267; S. Das Sarma, C. Nayak, S. Tewari, {\it Phys. Rev. B }{\bf 73} (2006)
220502; S. Tewari, C. Zhang, S. Das Sarma, C. Nayak, D. H. Lee, {\it Phys.
Rev. Lett. }{\bf 100} (2008) 027001.

\bibitem{optical1}  N. R. Cooper, N. K. Wilkin, J. M. F. Gunn, {\it Phys.
Rev. Lett. }{\bf 87} (2001) 120405; V. Gurarie, L. Radzihovsky, A. V.
Andreev, {\it Phys. Rev. Lett. }{\bf 94} (2005) 230403; S. Tewari, S. Das
Sarma, C. Nayak, C. Zhang, P. Zoller, {\it Phys. Rev. Lett. }{\bf 98} (2007)
010506.

\bibitem{ioffe}  L. B. Ioffe, V. B. Geshkenbein, M. V. Feigelman, A. L.
Fauch\`{e}re, G. Blatter, {\it Nature }{\bf 398} (1999) 679; G. Blatter, V.
B. Geshkenbein, L. B. Ioffe, {\it Phys. Rev. B }{\bf 63} (2001) 174511; L.
B. Ioffe, M. V. Feigelman, {\it Phys. Rev. B }{\bf 66} (2002) 224503; B.
Doucot, J. Vidal, {\it Phys. Rev. Lett. }{\bf 88} (2002) 227005; B. Doucot,
M. V. Feigelman, L. B. Ioffe, {\it Phys. Rev. Lett. }{\bf 90} (2003) 107003;
L. B. Ioffe, M. V. Feigelman, A. S. Ioselevich, D. Ivanov, M. Troyer, G.
Blatter, {\it Nature }{\bf 415} (2002) 503.

\bibitem{ioffe1}  B. Doucot, L. B. Ioffe, J. Vidal, {\it Phys. Rev. B }{\bf %
69} (2004) 214501; B. Doucot, M. V. Feigelman, L. B. Ioffe, A. S.
Ioselevich, {\it Phys. Rev. B }{\bf 71} (2005) 024505; B. Doucot, L. B.
Ioffe, {\it Phys. Rev. A }{\bf 72} (2005) 032303; B. Doucot, L. B. Ioffe,
{\it New J. Phys. }{\bf 7} (2005) 187; B. Doucot, L. B. Ioffe, {\it Phys.
Rev. B }{\bf 76} (2007) 214507.

\bibitem{witten1}  S. Elitzur, G. Moore, A. Schwimmer, N. Seiberg, {\it %
Nucl. Phys. B }{\bf 326} (1989) 108; G. Moore, N. Seiberg, {\it Comm. Math.
Phys. }{\bf 123} (1989) 177; E. Witten, {\it Comm. Math. Phys. }{\bf 121}
(1989) 351.

\bibitem{noi1}  G. Cristofano, V. Marotta, A. Naddeo, {\it Phys. Lett. B}%
{\bf \ 571} (2003) 250.

\bibitem{noi2}  G. Cristofano, V. Marotta, A. Naddeo, {\it Nucl. Phys. B}%
{\bf \ 679 }(2004) 621.

\bibitem{noi5}  G. Cristofano, V. Marotta, A. Naddeo, G. Niccoli, {\it J.
Stat. Mech.: Theor. Exper. }(2006) L05002.

\bibitem{noi}  G. Cristofano, V. Marotta, P. Minnhagen, A. Naddeo, G.
Niccoli, {\it J. Stat. Mech.: Theor. Exper. }(2006) P11009.

\bibitem{qcrew1}  A. Ekert, R. Jozsa, {\it Rev. Mod. Phys. }{\bf 68 }(1996)
733; A. Steane, {\it Rep. Prog. Phys. }{\bf 61 }(1998) 117; A. Galindo, M.
A. Martin-Delgado, {\it Rev. Mod. Phys. }{\bf 74 }(2002) 347.

\bibitem{ffxy1}  B. I. Halperin, D. R. Nelson, {\it J. Low Temp. Phys. }{\bf %
36 }(1979) 1165; J. Villain, {\it J. Phys. C: Solid State Phys. }{\bf 10 }%
(1977) 1717; T. C. Halsey, {\it J. Phys. C: Solid State Phys. }{\bf 18 }%
(1985) 2437; S. E. Korshunov, G. V. Uimin, {\it J. Stat. Phys. }{\bf 43 }%
(1986) 1.

\bibitem{foda}  O. Foda, {\it Nucl. Phys. B}{\bf \ 300 }(1988) 611.

\bibitem{orlando1}  T. P. Orlando, J. E. Mooij, L. Tian, C. H. van der
Wal,L. S. Levitov, S. Lloyd, J. J. Mazo, {\it Phys. Rev. B }{\bf 60 }(1999)
15398; J. R. Friedman, V. Patel, W. Chen, S. K. Tolpygo, J. E. Lukens, {\it %
Nature }{\bf 406 }(2000) 43.

\bibitem{fluct1}  Proceedings of the NATO Advanced Research Workshop on {\it %
Coherence in superconduting networks}, J. E. Mooij and G. Schon (Eds.), {\it %
Physica B} {\bf 152} (1988); Proceedings of the Conference on {\it %
Macroscopic quantum phenomena and coherence in superconducting networks}, C.
Giovannella and M. Tinkham (Eds.), World Scientific, Singapore, 1995; Y. M.
Blanter, R. Fazio, G. Schon, {\it Nucl. Phys. B }{\bf S58 }(1997) 79.

\bibitem{si1}  S. Doniach, {\it Phys. Rev. B }{\bf 24} (1981) 5063; L. J.
Geerligs, M. Peeters, L. E. M. de Groot, A. Verbruggen, J. E. Mooij, {\it %
Phys. Rev. Lett. }{\bf 63 }(1989) 326; R. Fazio, G. Schon, {\it Phys. Rev. B
}{\bf 43} (1991) 5307; H. S. J. van der Zant, L. J. Geerlings, J. E. Mooij,
{\it Europhys. Lett. }{\bf 19} (1992) 541; C. D. Chen, P. Delsing, D. B.
Haviland, Y. Harada, T. Claeson, {\it Phys. Rev. B }{\bf 50} (1995) 3959.

\bibitem{teitel}  S. Teitel, C. Jayaprakash, {\it Phys. Rev. Lett. }{\bf 51}
(1983) 1999; S. Teitel, C. Jayaprakash, {\it Phys. Rev. B }{\bf 27} (1983)
598.

\bibitem{halsey1}  T. C. Halsey, {\it Phys. Rev. B }{\bf 31} (1985) 5728.

\bibitem{stroud}  R. S. Fishman, D. Stroud, {\it Phys. Rev. B }{\bf 37}
(1987) 1499.

\bibitem{villain2}  J. Villain, {\it J. de Physique }{\bf 36 }(1975) 581; J.
Jos\'{e}, L. P. Kadanoff, S. Kirkpatrick, P. R. Nelson, {\it Phys. Rev. B }%
{\bf 16 }(1977) 1217.

\bibitem{fradkin}  E. Fradkin, B. A. Huberman, S. H. Shenker, {\it Phys.
Rev. B }{\bf 18 }(1978) 4789.

\bibitem{ktou}  J. M. Kosterlitz, D. J. Thouless, {\it J. Phys. C }{\bf 6 }%
(1973) 1181; J. M. Kosterlitz, {\it J. Phys. C }{\bf 7 }(1974) 1046.

\bibitem{granato1}  E. Granato, J. M. Kosterlitz, {\it Phys. Rev. B }{\bf 33}
(1986) 4767.

\bibitem{granato}  E. Granato, J. M. Kosterlitz, J. Lee, M. P. Nightingale,
{\it Phys. Rev. Lett. }{\bf 66} (1991) 1090; J. Lee, E. Granato, J. M.
Kosterlitz, {\it Phys. Rev. B }{\bf 44} (1991) 4819; E. Granato, J. M.
Kosterlitz, M. P. Nightingale, {\it Physica B }{\bf 222} (1996) 266.

\bibitem{gerardo}  G. Cristofano, G. Maiella, R. Musto, F. Nicodemi, {\it %
Nucl. Phys. B }{\bf 33C} {\it Proc. Suppl. (1993) }119.

\bibitem{theta}  D. Mumford, P. Norman, M. Nori, {\it Tata Lectures on Theta
III}, Birkhauser-Verlag, Boston (2006).

\bibitem{hall1}  R. B. Laughlin, {\it Phys. Rev. B }{\bf 23} (1981) 5632; R.
B. Laughlin, {\it Phys. Rev. Lett. }{\bf 50} (1983) 1395; R. B. Laughlin,
{\it Rev. Mod. Phys. }{\bf 71} (1999) 863.

\bibitem{Napoli-91-92}  G. Cristofano, G. Maiella, R. Musto, N. Nicodemi,
{\it Mod. Phys. Lett. A} {\bf 7 } (1992) 2583.

\bibitem{exper}  C. J. Lobb, {\it Physica }{\bf B126 }(1984) 319; D. W.
Abraham, C. J. Lobb, M. Tinkham, T. M. Klapwijk, {\it Phys. Rev. B }{\bf 26}
(1982) 5268; R. F. Voss, R. A. Webb, {\it Phys. Rev. B }{\bf 25} (1982)
3446; C. J. Lobb, D. W. Abraham, M. Tinkham, {\it Phys. Rev. B }{\bf 27}
(1983) 150; B. J. van Wees, H. S. J. van der Zant, J. E. Mooij, {\it Phys.
Rev. B }{\bf 35 }(1987) 7291; H. S. van der Zant, W. J. Elion, L. J.
Geerligs, J. E. Mooij, {\it Phys. Rev. B }{\bf 54 }(1996) 10081; P.
Martinoli, C. Leemann, {\it J. Low Temp. Phys. }{\bf 118 }(2000) 699.

\bibitem{rogalla}  H. Hilgenkamp, Ariando, H. J. H. Smilde, D. H. A. Blank,
G. Rijnders, H. Rogalla, J. R. Kirtley, C. C. Tsuei, {\it Nature} {\bf 422}
(2003) 50; J. R. Kirtley, C. C. Tsuei, Ariando, H. J. H. Smilde, H.
Hilgenkamp, {\it Phys. Rev. B }{\bf 72} (2005) 214521.

\bibitem{pi}  A. I. Larkin, Yu. N. Ovchinnikov, {\it Sov. Phys. Rev. JETP }%
{\bf 20 }(1965) 762; P. Fulde, R. A. Ferrel, {\it Phys. Rev. }{\bf 135 }%
(1964) A550.

\bibitem{kirtley}  J. R. Kirtley, C. C. Tsuei, {\it Rev. Mod. Phys. }{\bf 72
}(2002) 969; A. A. Golubov, M. Yu. Kuprianov, E. Il'ichev, {\it Rev. Mod.
Phys. }{\bf 76 }(2004) .

\bibitem{sfs}  V. V. Ryazanov, V. A. Oboznov, A. Y. Rusanov, A. V.
Veretennikov, A. A. Golubov, J. Aarts, {\it Phys. Rev. Lett. }{\bf 86 }%
(2001) 2427; V. V. Ryazanov, V. A. Oboznov, A. V. Veretennikov, A. Y.
Rusanov, {\it Phys. Rev. B} {\bf 65} (2002) 020501; M. Weides, M. Kemmler,
H. Kohlstedt, R. Waser, D. Koelle, R. Kleiner, E. Goldobin, {\it Phys. Rev.
Lett. }{\bf 97} (2006) 247001.

\bibitem{sifs}  T. Kontos, M. Aprili, J. Lesueur, F. Genet, B. Stephanidis,
R. Boursier, {\it Phys. Rev. Lett. }{\bf 89} (2002) 137007; M. Weides, M.
Kemmler, E. Goldobin, D. Koelle, R. Kleiner, H. Kohlstedt, {\it App. Phys.
Lett. }{\bf 89} (2006) 122511.

\bibitem{tunnel1}  O. Vavra, S. Gazi, D. S. Golubovic, I. Vavra, J. Derer,
J. Verbeeck, {\it Phys. Rev. B} {\bf 74} (2006) 020502.

\bibitem{tunnel2}  J. P. Cleuziou, W. Wernsdorfer, V. Bouchiat, T.
Ondarcuhu, M. Monthioux, {\it Nature Nanotech.} {\bf 1} (2006) 53.

\bibitem{tunnel3}  J. A. van Dam, Y. V. Nazarov, E. P. A. M. Bakkers, S. De
Franceschi, L. Kouwenhoven, {\it Nature }{\bf 442} (2006) 667.

\bibitem{cft}  P. Di Francesco, P. Mathieu and D. Senechal, {\it \ Conformal
Field Theories}, Springer-Verlag, Berlin (1996).
\end{thebibliography}
\end{document}